# Empirical analysis of the ship-transport network of China

Xinping Xu [1,*], Junhui Hu [2], Feng Liu [1]

[1] *Institute of Particle Physics, HuaZhong Normal University, Wuhan 430079, China*
[3] *College of physics and electronic engineering, Guangxi Normal University, Guilin 541004, China*

*Abstract*: Structural properties of the ship-transport network of China (STNC) are studied in the light of recent investigations of complex networks. STNC is composed of a set of routes and ports located along the sea or river. Network properties including the degree distribution, degree correlations, clustering, shortest path length, centrality and betweenness are studied in different definition of network topology. It is found that geographical constraint plays an important role in the network topology of STNC. We also study the traffic flow of STNC based on the weighted network representation, and demonstrate the weight distribution can be described by power law or exponential function depending on the assumed definition of network topology. Other features related to STNC are also investigated.

**Leading paragraph:**

**Transportation systems are of great importance to the development of a country and are important indicators of its economic growth. They form the backbone of tourism industry, support movement of goods and people across the country, thereby driving the national economy. During the past few years, complex network analysis has been used to study transportation systems. In this paper, we study the statistical properties of the ship-transport networks of China (STNC)in which the nodes are ports and the links are passenger liners connecting the ports. Based on two different representation of network topology －the space L and space P, where in the space L two ports are considered to be connected if they are consecutive stops on a route of at least one ship while in the space P two ports are connected when there is at least one ship which stops at both ports, we explore scaling laws and correlations that may govern intrinsic features of STNC. Our results show that the degree distribution in the space L follows a power law whereas in the space P exhibits a truncated scale-free behavior. Both topologies display the small world properties, but the large average clustering coefficient and small average path length in the space P indicates that the small-world property is more visible in the space P. We also study different kinds of centrality measures as an indicator of the importance of a specific port, and find strong correlations**

**between different kinds of centralities. Finally, we study the traffic flow for the weighted version of STNC, and power law behaviors are also observed for the traffic distribution and strong correlations between the strengths are investigated in both the network topology. Comparing to other transport networks, STNC has a combined structure of the airport network and railway transport network (or bus-transport network). Therefore, we conjecture that STNC has an intermediate topology between airport network and the railway network.**

## 1. Introduction

During the past few years, several spatial networks in which the nodes occupy a precise position in two or three-dimensional Euclidean space and edges are real physical connections have received much attention from the complex network point of view.[1-3] Typical examples include neural networks,[4] information/communication networks,[5,6] electric power grids,[7] and transportation systems ranging from river,[8] airport,[9-13] street,[14] bus,[15,16] railway,[17,18] subway,[19-20] and ant networks of galleries[21] etc. Most of the works in the literature have focused on the characterization of the topological properties of spatial networks, including the small-world behavior and scale free structure.[22,23]

In this paper, we present an investigation of the ship-transport network of China (STNC). Previous analysis on ship-transport network can trace back to the work of Pitts in the last century.[8] Pitts studied the medieval Russian river network, in an effort to assess the centrality of the urban places on the graphs, and showed that Moscow was most central and accessible with aggregate least effort.[8] Pitts's work had received much attention by historians and geographers concerning the growth of Moscow. Here, we consider the ship-transport network of China, which comprises 42 sea ports and 120 river ports in different locations. We have gathered the ship schedule information from the Internet.[24] The nodes of the network are the ports and the edges are the lines connecting them along the route. The results presented below are based on a large number of passenger liners, which only carries passengers and the cargo transport are not considered in our analysis. Compared with other types of transport networks, the ship-transport network has a more complex structure due to the following features: 1) the topological structure is constrained by their geographical embedding. Since the ports are distributed along the sea or river and the ship can only travel along the river, this imposes a strong constraint to the structure of STNC. In addition, the existence of branching rivers also makes its topological structure significant. 2) Directed, weighted links with slightly fluctuating frequency. Quite similar to the airport networks,[13] there may be many direct ship schedules connecting two neighboring ports A and B every day. Therefore asymmetric matrices are used to characterize properties (in-degrees, out-degrees, etc) which are sensitive to the direction. 3) Some ship schedules travel through more than two ports. Therefore, there are two topological representations for STNC, the so-called space L and space P. Based on the concepts of space L and P, we will show that scaling laws may govern intrinsic features of the physical quantities. This gives analogues feature to the railway network or bus-transport network.[15,17]

The paper is organized as follows: In the next section, we give a brief description of the concepts of space L and space P. In section 3, we use some matrices to denote the structure of STNC in different definition of network topology. Section 4 presents the results on degree distributions and degree correlations of STNC. Section 5 analyzes the clustering and path length. Section 6 is devoted to the study of centrality measures and betweenness. In section 7, we will explore the traffic flow of weighted STNC. Discussion and conclusions are given in the last part,

Sec. 8.

**2. The concepts of space L and P**

Sen et al. [17,15] have introduced a new topology describing the public transport networks – the idea of the space P which indicating two arbitrary stations (or ports in this paper) are connected by a link when there is at least one vehicle (ship) which stops at both the stations (ports). Generally, we can present the public transport network in two different topological representations. The first topological representation is the space L which consists of nodes being ports, and a link between two nodes exists if they are consecutive stops on the ship route. The node degree k in this topology is just the number of different ship routes one can take from a given ports. The distance in such a space is measured by the total number of stops passed on the shortest path between two nodes. The second representation is the space P, in which an edge is formed between two nodes given that there is a ship schedule traveling between them. Consequently, the node degree $k$ in this topology is the total number of nodes reachable using a single ship route and the distance can be interpreted as a number of transfers (plus one) one has to take to get from one port to another. It is obvious that in the space P the distances are numbers of transfers (plus one) needed during the travel and distances are much shorter than in the space L. We point out that the spaces $L$ and $P$ do not take into account the geographical distance between nodes. Such a method is analogues to the description of several other types of networks: Internet,[25] power grids,[26] etc. Both spaces are presented at Fig1 (a) and (b).

We extend the idea of space L and P to the case of directed network.[15] See Fig.1, suppose line A and line B consist of 5 consecutive ports starting from port 1 to port 5 and from port 6 to port 9 with a crossing port 3 in the two lines, the directed representation of space L and P are illustrated in Fig1 (c) and (d).

## 3. Network representation

Structure of STNC can be symbolized by an asymmetrical weight matrix $W$ whose element $W_{ij}$ is the number of passenger liners traveling from port $i$ to port $j$. In this perspective, we can use two different matrices $W^L$ and $W^P$ to denote the traffic flow of STNC in the space L and P respectively. We should note that the element $W_{ij}$ includes contribution from the direct ship transportation between $i$ and $j$ without middle stops. As a matter of fact, the direct ship between two ports gives the same contribution to the weights in the space L and space P. Based on the directed representation of space L and P, we define a set of quantities to characterize the properties of STNC.

First, we employ $k_{in}^L(i)$ and $k_{ou}^L(i)$ to denote the in-degree and out-degree of a given node $i$ for directed STNC in the space L, and $k_{un}^L(i)$ to represent the undirected degree of the undirected STNC in the same network topology. The in-degree and out-degree of port i stand for number of ports from which can arrive at port i and the number of ports can be reached from port i, respectively. Using the weighted matrix, we can write these quantities as follows

$$k_{ou}^L(i) = \sum_{j \neq i} h(W_{ij}^L - 1) \tag{1}$$

$$k_{in}^L(i) = \sum_{j \neq i} h(W_{ji}^L - 1) \tag{2}$$

And
$$k_{un}^L(i) = \sum_{j \neq i} h(W_{ij}^L + W_{ji}^L - 1) \tag{3}$$

where $h(x)$ is a unit step function, which takes 1 for $x \geq 0$ and 0 otherwise. Similarly, we can also obtain the above equations for the space P. As STNC is a directed connected network with 162 nodes, we report the size of the giant strongly component, in-component and out-component defined in Ref.[27]. We find that the strongly connected component, i.e., every pair of ports is connected in both directions, comprises 154 ports. The size of in-component and out-component are found to be 160 and 158, respectively. This indicates that the corresponding adjacency matrix for the network is almost symmetrical.

Next, we consider the traffic flow of weighted STNC. We define the total traffic flow coming into node $i$ as the in-strength $s_{in}^L(i)$ and use $s_{ou}^L(i)$ to denote the out-strength in the space L. Therefore,

$$s_{ou}^L(i) = \sum_{j \neq i} W_{ij}^L \qquad (4)$$

$$s_{in}^L(i) = \sum_{j \neq i} W_{ji}^L \qquad (5)$$

And the total strength is
$$s_{total}^L(i) = \sum_{j \neq i} (W_{ji}^L + W_{ij}^L) \qquad (6)$$

Eq.(4),(5) and (6) also hold for the space P. As mentioned in the introduction, STNC consists of 162 ports, the small network size gives some influence to the statistical significance of our results. The total number of traffic of the space L is 11480, which means that there are 11480 ships (links) connecting the ports. Compared to the space L, the total traffic of the space P has a much large value, 61060. The large amount of traffic in both spaces suggests that database is highly redundant in its topological structure, i.e., most connections between pairs of ports are represented by more than one ship. This makes the analysis of network topology reliable.

## 4. Degree distributions and degree correlations

In order to reduce the statistical errors arising from the limited system size, we use the cumulative distribution to describe the relevant variables. The cumulative degree distributions $P_{cum}(k_{in})$, $P_{cum}(k_{ou})$ and $P_{cum}(k_{un})$ in the space L and P are shown in Fig. 2. In Fig.2 (a), as we can see, all the three degree distribution in the space L can be approximately described by the same power law $P_{cum}(k) \sim k^{-g}$ with $g = 1.7$. The exponent of the original degree distribution $g + 1 = 2.7$ is quite close to the exponent in the BA model of evolving networks with preferential attachment. The power law behavior of degree distribution in the space L is quite similar to that of the bus-transport network, where the degree in the space L also follows power law decay.[15] The three cumulative degree distributions in the space P is plotted in Fig. 2 (b). It is amazing to find that all the three distributions follow nearly the same truncated power law decay with a cutoff value $k \sim 20$. The distributions decay slowly for small $k^P$ while decay drastically for large $k^P$. This is quite similar to the case of Airport Network of China (ANC),[13] which has a two-regime power-law degree distribution with a turning point $k_c \sim 26$.

Now, we explore an important feature of STNC, the degree correlations. First we consider the correlation between in-degrees and out-degrees in the space L and P. Fig.3 (a) is a plot of out-degree $k_{ou}$ versus in-degree $k_{in}$. We find that $k_{ou} \approx k_{in}$ since most of the ship routes are bidirectional. In succession, we study the degree-degree correlations for undirected STNC. Fig.3 (b) shows the average nearest neighbor degree $<k^{nn}>$ as a function of the degree k in the space L and P. It is found that $<k^{nn}>$ is a decreasing function of k in the space L while it is an increasing function of k in the space P. We also check the degree correlation coefficient *r* which is a global quantitative measure of degree correlations defined by Newman,[28] and find its value in space L and P to be -0.18 and 0.27. This indicates STNC shows disassortative mixing in the space L but assortative mixing in the space P, which is consistent with the observation in Fig.3 (b). One possible explanation for this strange behavior is that it arises as a result of the topology difference. In order to present this, we define the *line length l* as the number of ports that the line has. Fig.3 (c) shows the line length distribution of STNC. As we can see, most of the ship routes are direct port to port connections (*l =2*), the topology of the space L approaches a star structure in which low degree nodes connects to high degree nodes. On the contrary, in the space P, routes with large line length cross each other in more than one port, resulting in lots of connections between large degree nodes.

This may lead to the transition of the degree mixing pattern observed in the space L and P. An interesting result related to the degree correlation is that small networks N <500 seem to be dissortative and large networks N > 500 appear to be assortative.[15] Here we demonstrate that this is not the general case as STNC also has a small network size N=162. Another kind of degree correlation is the correlation of the degrees between the space L and space P as shown in Fig3 (d). Most of the points is beyond the bisector of the coordinate, i.e., $k_{un}^P \geq k_{un}^L$, which is consistent with our intuition that degrees in the space P is equal to or larger than that of the space L. Other parameters characterizing STNC are summarized in Table I.

**5. Clustering and path length**

In this section, we consider the clustering and the shortest path length of undirected STNC. First, we check the clustering spectrum, or the degree-dependent local clustering, which can be described by the clustering coefficient as a function of degree. In many instances, the clustering spectrum exhibits also a power-law behavior, $C(k) \sim k^{-a}$. A value of $a$ close to 1 has been empirically observed in several real networks, and analytically found in some growing network models.[29] The clustering spectrum in the space L and P is shown in Fig.4. As seen in Fig.4 (a), the degree dependent clustering in the space L can be described by a power law function $<c(k)> \sim k^{-1}$ with average clustering coefficient $<c>=0.54$. This nontrivial scaling of clustering may indicate a hierarchical and modular structure of STNC.[29,30] Interestingly, the clustering spectrum in the space P shows a peak around 15. The average clustering coefficient in the space P is found to be 0.83. The clustering coefficient in both the network topology has a large value compared to the corresponding value for classical random graph with the same average degree, in contrast to the public transport networks[15] which have a small clustering in the space L and large clustering in the space P.

Fig.4 (b) is a plot of distribution of the shortest path length between two ports. Ranges of distances in the space L are much broader as compared to corresponding ranges in the space P which is a natural effect of topology differences. The shape of path length distribution can be explained in the following way: because transport networks tend to have an inhomogeneous structure, it is obvious that distances between ports lying on the branching routes are quite large and such a behavior gives the effect of observed long tails in the distribution. On the other hand shortest distances between ports belonging to the main rivers are more random and they follow the Gaussian distribution. A combined distribution has an asymmetric shape with a long tail for large paths.

The average shortest path length in the space L and P are measured to be 5.86 and 3.87 respectively. The characteristic length 4 in the space P means that in order to travel between two different ports one needs in average no more than three transfers. Other transport networks also share this property, depending on the system size. The large average path length in the space L and P suggests that the network topology is strongly constrained by geographical location of the ports. In order to address this issue, we calculate the average path length for the randomized network with the same degree sequence. This method has been extensively used in contraposition to real networks with the same degree distribution. The average path lengths of the randomized network in the space L and P are found to be 3.20 and 2.20, which is much smaller than the counterpart of STNC. The ports are separated by the Euclidean distance, thus their ability connecting to other far ports is restricted. For instance, there is no direct scheduled ship for the most connected two ports Shanghai and HongKong, although they connects to 18 and 21 other ports respectively. The reason is simple: the establishment of connection between the two ports will cost too much since the distance is far. Therefore the geographical constraint is strong in the case of STNC.

We need to stress that the large value of clustering coefficient and small value of average path length in the space P, indicates the small-world properties are more visible in such network topology. From the view of passengers, to reach from any port to any other port in the network, one should make fewer changes of ships as possible. However, it is not economically feasible for the ship service providers. We conclude that the ship-transport network has evolved with the efforts to satisfy both the convenience of passengers and the benefits of ship service providers, thus endow STNC the small-world topology.

**6. Centrality and Betweenness**

Although the degree of a node is a centrality measure describing the "importance" of a specific node, it does not provide complete information for understanding the structural properties of transport networks. To start to address this issue, we use various measures of structural centrality to quantify the importance of a port. The first kind of centrality of a node is the total strength $S_{total}$, which will be studied in the next section. To shed some light on the relationship between the strength and degree, we plot the average total strength $<S_{total}>$ as a function of the degree k in Fig.5 (a). As we can see, in the space P, the relation can be well approximated by a power-law dependence $<S_{total}> \sim k^b$ with an exponent $\beta = 2.0 \pm 0.1$. This implies that the strength of nodes grows faster than their degree. On the contrary, in the space L, $<S_{total}>$ increase slowly for large k. This gives similar results in Ref.[12]. The different behavior of $<S_{total}>$ in the space L and P may be attributed to the topology differences. The second kind of centrality measure is $D_i$, which is defined as the average shortest path length from a certain vertex i to all other vertices as follows,

$$D_i = (\sum_{j \neq i} L_{ij})/(N-1) \qquad (7)$$

Where $L_{ij}$ is shortest path length between port i and j. This quantity is a centrality measurement of node i, the small value of $D_i$ suggests that it is convenient from node i to other nodes. Fig.5 (b) shows $<D(k)>$ as a function of degree *k*. The straight line in the linear-log plot indicates the relation $<D(k)>=A-Blog(k),$ which can be proved using a simple model of random graphs and generating function formalism or a branching tree approach.[31,32] The above relation exists in both the network topology and is a universal scaling for complex networks.

The third kind of centrality is betweenness. The betweenness $B_i$ of port *i* is defined as the number of shortest paths connecting any two ports that involve a transfer at port *i*.[33-35] We define the normalized betweenness as $b_i=B_i/<B>$, where $<B>$ is the average betweenness for the network. We demonstrate the cumulative distribution $P_{cum}(b)$ of the normalized betweenness in Fig.5 (c). It is surprising to find that the distribution in the space L can be described by a power-law $P_{cum}(b) \sim b^{-1.2}$ while distribution in the space P follows a two-regime power law with two different exponents, known as double Pareto law,[36] with a turning point at betweenness value $b_c$, which can be well prescribed by the following expression,

$$P_{cum}(b) \sim \begin{cases} b^{-\gamma_1}, \text{for} \quad b \leq b_c \\ b^{-\gamma_2}, \text{for} \quad b > b_c \end{cases} \quad (5)$$

where $g_1$ and $g_2$ are the respective exponents of two separate power laws. By means of fitting, exponents pairs $(g_1, g_2)$ of the cumulative distribution is equal to 0.5 and 2 respectively. The critical value $b_c$ is found to be 3. The truncated cumulative distribution of normalized betweenness can be ascribed the truncated degree distribution. In order to present this, we show the average normalized betweenness $<b>$ as a function of degree k in Fig.5 (d). As we can see, it is a power-law growth $<b(k)> \sim k^\eta$ with exponent $\eta = 1.0 \pm 0.1$ for space L and $\eta = 2.2 \pm 0.1$ for space P, although there are some deviations for small $k^L$ and large $k^P$. Such a behavior indicates that the degree of a node and its betweenness centrality are strongly correlated, i.e., highly connected nodes are also the most central ports. The strong correlation between <b> and k gives a possible explanation to the observed truncated betweenness distribution.

**7. Traffic dynamics of weighted STNC**

To include the information about the amount of traffic flow on STNC, we analyzed the weighted STNC, by considering the flow of information (traffic) on the topology of the network.[3] As mentioned in section 3, the weighted STNC can be described by a weight matrix, W, where each element $W_{ij}$ stands for the total number of ships traveling from port *i* to port *j*. Since majority of the ship schedules are bidirectional in STNC, traffic flow between two port *i* and *j* are symmetric, *i.e.*, $W_{ij} \approx W_{ji}$. The statistical analysis of weights between pairs of nodes indicates the presence of right-skewed distribution as shown in Fig. 6 (a) and (b). Fig6 (a) is a log-log plot of the cumulative weight distribution in the space L. The distribution exhibits a power law behavior in the region (10~100). This shows a high level of heterogeneity in STNC, as also found in the case of Airport Network of China[13] and WAN[12]. The cumulative weight distribution in the space P is shown in Fig6 (b). The data follows an exponential decay. Such a decay function indicates that the probability of finding a very busy port-pair is nonzero and significant instead. Because the weight of two ports in the space P is larger than that in the space L, this provides a possible explanation to different distribution in the space L and P.

Since the individual weights do not provide a general picture of the network's complexity, we investigate the total strength $s_{total}$ of each port. We find that the cumulative strength distribution of STNC follows a power law as seen in Fig. 6 (c) and (d) for a wide range of s, although it deviates from it for large strengths. The two distributions are power law decay function in small strength while with a sharp cutoff in large values of strength. The deviation for large strengths can be attributed to the cost of adding links to nodes or to the limited capacity of nodes. With the help of numerical simulations, it was shown[9] that the cost of adding links leads to a cutoff of the power-law regime for large strength, as is the case of STNC.

In order to reveal the correlations between the strength, we plot the in-strength $s_{in}$ versus out-strength $s_{ou}$ in Fig.7 (a) and (b). As we can see, the in-strength $s_{in}$ is close to the out-strength $s_{ou}$ in both the network topology. This is quite natural for STNC because each port should generally maintain the balance of its traffic flow. In Fig7(c), we give the relation of the total strength in the space L and P. For most of the ports, the total strength of space P is larger than that of the space L, and only for a small number of ports $s_{total}^{P} = s_{total}^{L}$. Since the direct ships between two ports gives equal contribution to the weights in both the space, and ship routes with more than two ports contribute more to the strength in the space P compared to the space L, we conclude that there are only direct ships connecting the ports on the bisector. The correlations among weighted quantities and the underlying topological structure provide us a better understanding for the hierarchies and organizational principles of complex weighted network.

TABLE I. Parameters of STNC in the space L and P. N is the number of ports, l is total number of ships between the ports, $<k_{un}>$ stands for the average degree, r is the degree correlation coefficient, $<C>$ is the average clustering coefficient and $<L>$ is the average path length of undirected STNC. $<W>$ is the average weight while $<S_{in}>$ and $<S_{ou}>$ are the average in-strength and out-strength.

| Parameter / Space | N | l | $<k_{un}>$ | r | $<C>$ | $<L>$ | $<W>$ | $<S_{in}>$ | $<S_{ou}>$ |
|---|---|---|---|---|---|---|---|---|---|
| Space L | 162 | 11480 | 3.11 | -0.18 | 0.54 | 5.86 | 25.06 | 70.87 | 70.86 |
| Space P | 162 | 61060 | 8.27 | 0.27 | 0.83 | 3.87 | 47.01 | 376.92 | 376.91 |

## 8. Conclusion and discussion.

In conclusion, we have analyzed the statistical properties of STNC. We explore scaling laws and correlations that may govern intrinsic features of such network. The topological properties, including the degree distribution, clustering, the shortest path length, centrality and betweenness are studied in the space L and P respectively. The degree distribution follows a power law or truncated power law depending on the network topology definition. The degrees are negatively correlated in the space L while positively correlated in the space P, which can be confirmed by the measured values of degree correlation coefficient. Small-world behavior is observed in both topologies but it is much more distinct in the space P and the hierarchical structure of network is also deduced from the behavior of $c(k)$. The relative large value of the average path length compared to the randomized version of STNC indicates that the geographical constraint is strong in the case of our network. Different measures of centrality, including degree $k$, strength $s$, average shortest path length from a certain vertex to all other vertices $D$ and betweenness $b$, are strongly correlated. Finally, we study the traffic dynamics of weighted STNC. The weight distribution follows a cutoff power law in the space L while the distribution in the space P is an exponential decay. Power law behaviors are also observed for the strength distribution and strong correlations between the strengths are investigated in both the network topology.

We would like to point out that, since STNC is a weighted, directed and dual-topology network, STNC has a combined structure of the airport network and railway transport network (or bus-transport network). We conjecture that STNC has an intermediate topology between airport network and the railway network. One can propose an evolution model for the ship-transport network with a simple idea which can be realized through computer simulation. At the beginning of the ship-transport development, there are a few ships connecting only the geographical neighboring ports. In the subsequent evolution, more ship routes are established to connect river ports and sea ports according to the spatially preferential attachment. The newly ships (or lines) are established by optimizing an objective function involving both the geographical and topological ingredients. Such combined ingredients may endow STNC the complex structure in both the network topology and traffic dynamics.

**8. Acknowledgement** This work is supported by NSFC under projects 10375025, 10275027 and by the MOE under project CFKSTIP-704035.

*Corresponding author. E-mail address: xuxp@ihep.ac.cn

Figure captions:

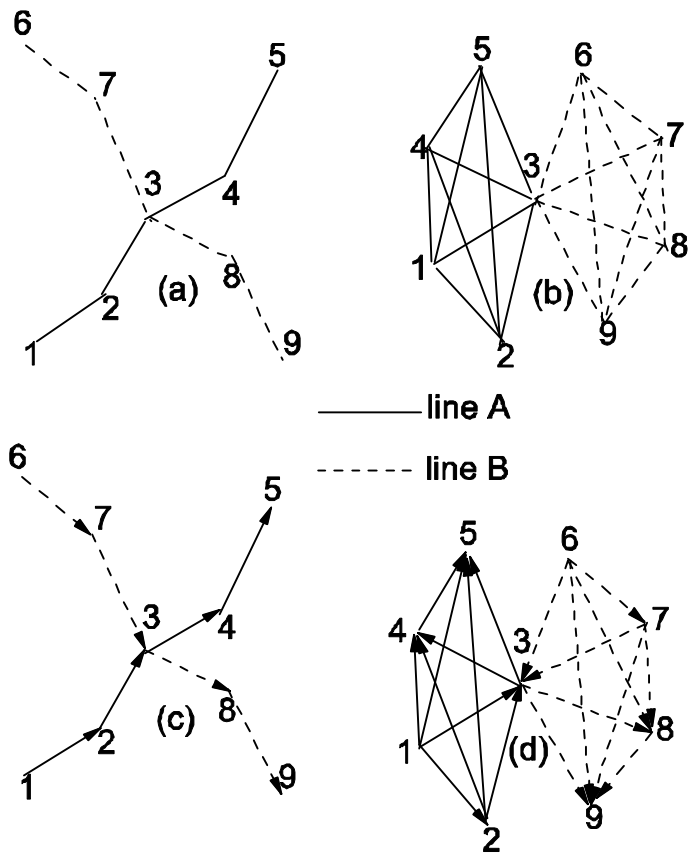

Fig.1 Illustration of the space L and the space P.   (a) and (b) are the undirected representation in the space L and space P. (c) and (d) are the directed representation corresponding to (a) and (b). The space L consists of nodes representing ports and links between two nodes exists if they are consecutive stops on the route. The node degree k in this topology is just the number of different ship routes one can take from a given ports. In the space P, the nodes are the same as in the space L, an edge between two nodes means that there is a ship schedule traveling between them.

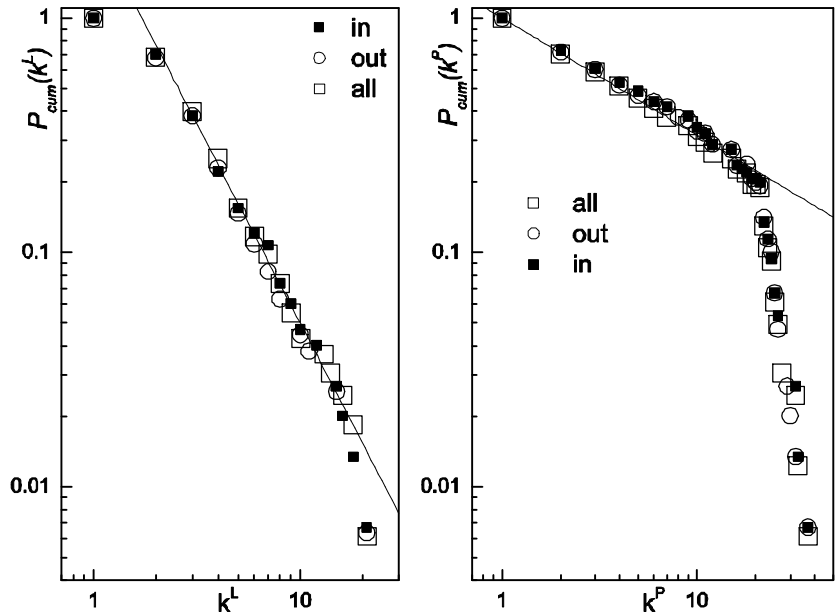

Fig.2. Cumulative degree distributions in the space L (a) and space P (b). The in-degree, out-degree and undirected degree distributions in the space L can be approximately described by the same power law $P_{cum}(k) \sim k^{-1.7}$. On the contrary, all the three distributions in the space P follow nearly the same truncated power law decay with a cutoff value $k \sim 20$.

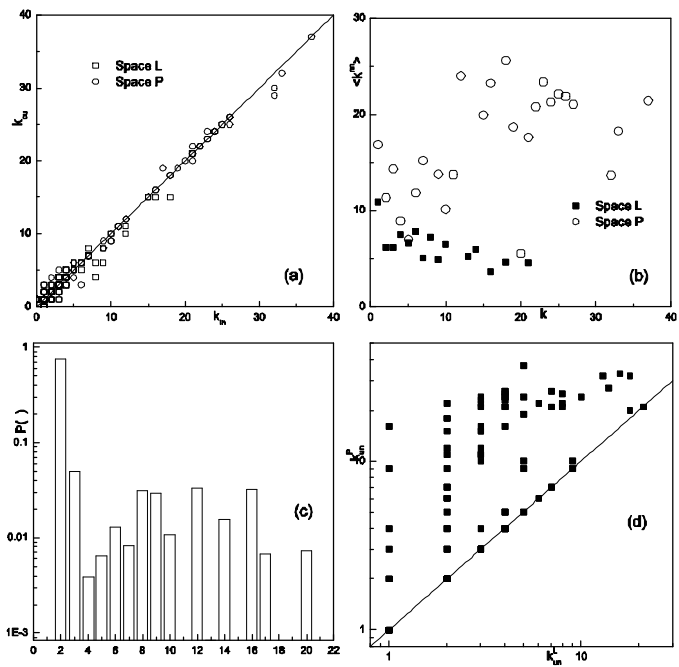

Fig.3. Degree correlations in the space L and space P. (a)In-degree versus out-degree. (b) Average nearest neighbor degree <k$^{nn}$> as a function of degree k. (c) Probability distribution of the line length λ. (d)Degrees of space P versus that of space L for undirected STNC.

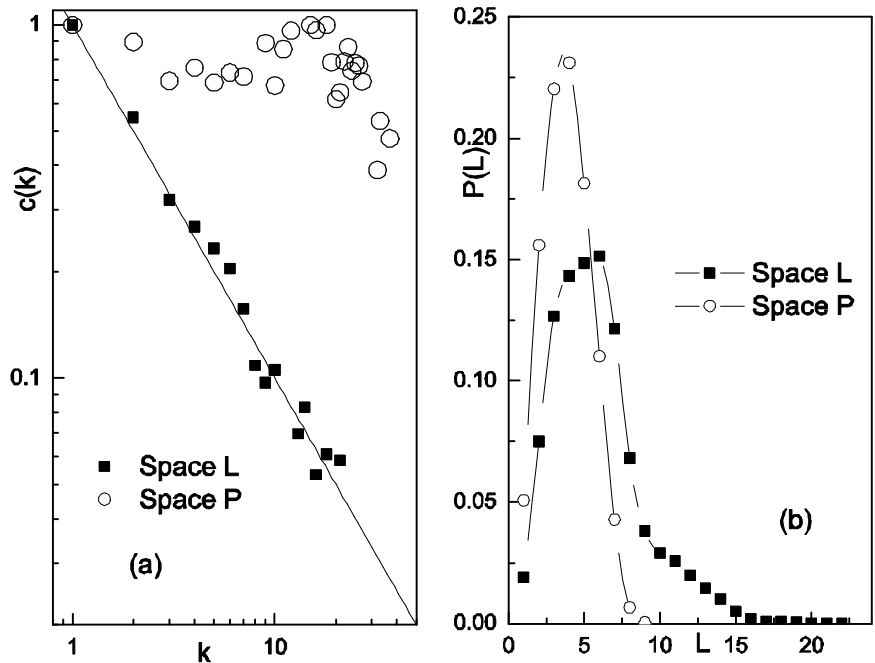

Fig.4. (a) Average local clustering <c(k)> as a function of the degree in two different spaces. As we can see, the degree dependent clustering in the space L shows a power law behavior *<c (k)> ~k$^{-1}$* with average clustering coefficient *<c>=0.54*. This nontrivial scaling of clustering may be an indicator of hierarchical and modular structure of STNC. Interestingly, the clustering spectrum in the space P shows a peak around 15. (b) Path length distributions in the two spaces. The distribution in the space L demonstrates a much broader range of distances compared to the case of space P.

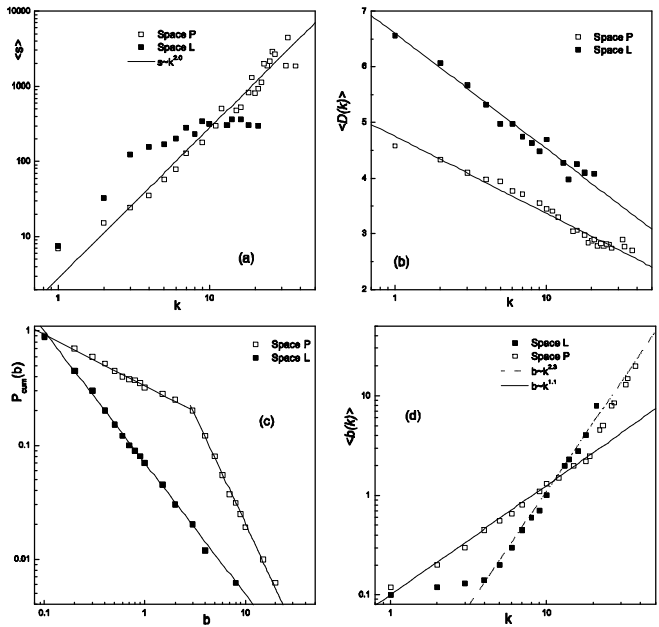

Fig.5 Different measures of centrality: degree *k*, average strength *<s(k)>*, average shortest path length from a certain node to all other nodes *<D(k)>* and betweenness *b*. (a) Average strength *<s(k)>* as a function of the degree *k* of nodes. (b) Average shortest path length from a certain node to all other nodes *<D(k)>* versus degree *k*. (c) Cumulative distribution of the normalized betweenness b. (d)Relationship between the average normalized betweenness*<b(k)>* versus degree *k*. As shown in the figure, different kinds of centrality measures are strongly correlated, i.e., the average strength *<s(k)>* and betweenness *<b(k)>* increase with degree *k* while average shortest path length from a certain node to all other nodes *<D(k)>* decreases with degree *k*.

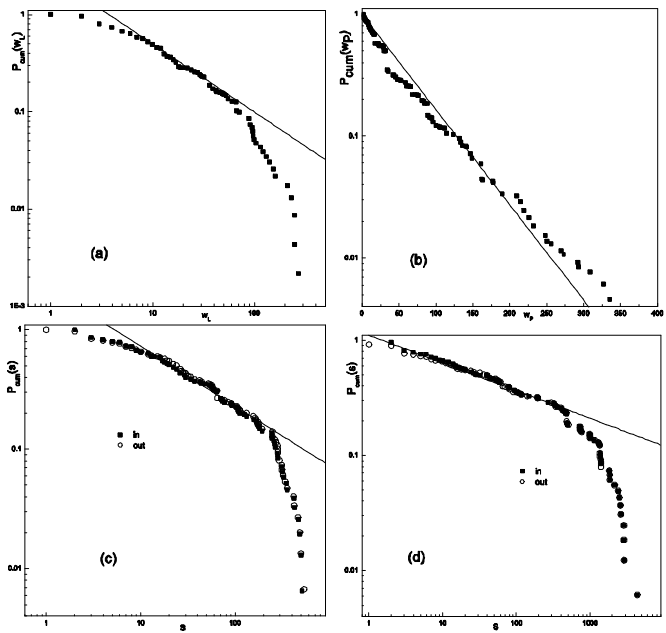

Fig.6 Cumulative weight distributions in the space L (a) and space P (b). The distribution in the space L exhibits a power law behavior in the region (10~100). In contrast, the distribution in the space P follows an exponential decay. The cumulative strength distributions in the space L and P are shown in (c) and (d). The distributions are power law decay function in small strength range while with a sharp cutoff in large values of strength as described in the text.

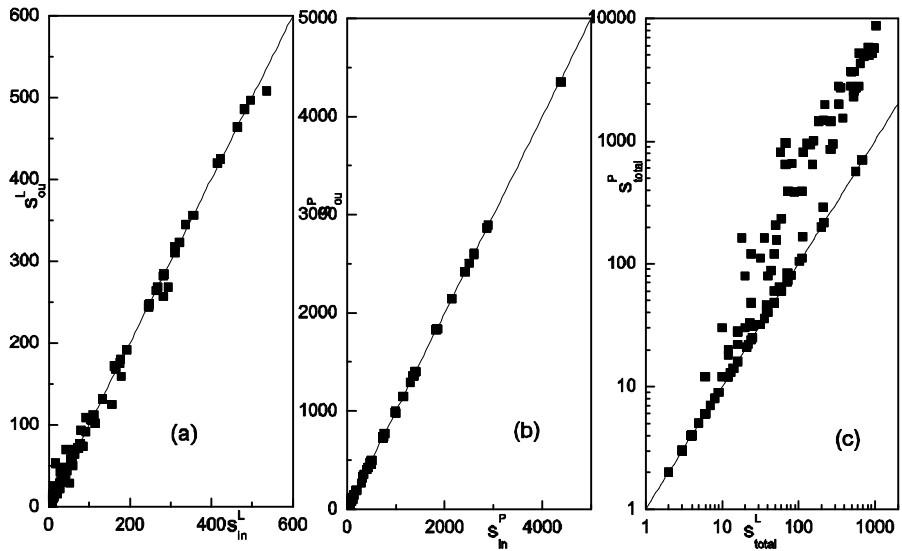

Fig.7 Correlations between strengths in two different spaces. (a) Out-strength versus in-strength in the space L, (b) The same plot as (a) but for the space P. (c) Total strength $s^P_{total}$ in the space P versus total strength $s^L_{total}$ in the space L. The solid lines are bisector of the coordinate axis. The relation $s_{ou} \approx s_{in}$ in both the spaces indicates that each port generally maintain the balance of its traffic flow.